\documentclass{article}

\usepackage{arxiv}

\usepackage[utf8]{inputenc} 
\usepackage[T1]{fontenc}    
\usepackage{hyperref}       
\usepackage{url}            
\usepackage{booktabs}       
\usepackage{amsfonts}       
\usepackage{nicefrac}       
\usepackage{microtype}      
\usepackage{lipsum}
\usepackage{graphicx}

\title{Balancing simulation and gameplay - Applying Games User Research to LeukemiaSIM}

\author{
 Erin Brintnell \\
  Department of Computer Science\\
  University of Calgary\\
  2500 University Dr NW\\
  Calgary, AB, T2N 1N4\\
  \texttt{erin.brintnell@ucalgary.ca} \\
   \And
 Owen Brierley \\
  Department of Computer Science\\
  University of Calgary\\
  2500 University Dr NW\\
  Calgary, AB, T2N 1N4\\
  \texttt{owen.brierley@ucalgary.ca} \\
  \And
 Neil Christensen \\
  Department of Computer Science\\
  University of Calgary\\
  2500 University Dr NW\\
  Calgary, AB, T2N 1N4\\
  \texttt{neil.christensen1@ucalgary.ca} \\
   \And
  Christian Jacob \\
  Department of Computer Science\\
  University of Calgary\\
  2500 University Dr NW\\
  Calgary, AB, T2N 1N4\\
  \texttt{cjacob@ucalgary.ca} \\
}

\begin{document}
\maketitle
\begin{abstract}
A bioinformatics researcher and a game design researcher walk into a lab... This paper shares two case-studies of a collaboration between a bioinformatics researcher who is developing a set of educational VR simulations for youth and a consultative game design researcher with a background in  Games User Research (GUR) techniques who assesses and iteratively improves the player experience in the simulations. By introducing games-based player engagement strategies, the two researchers improve the (re)playability of these VR simulations to encourage greater player engagement and retention.
\end{abstract}

\keywords{games user research, patient education, virtual reality}

\section{Introduction}
\textit{"Have you heard the one about a bioinformatics researcher and a game design researcher walking into a lab?"} Joking aside, this rare situation actually happened. The bioinformatics researcher brought a virtual reality (VR) experience they have been working on for the pediatric clinical setting. The game design researcher played the game and observed a dichotomy between a strong educational foundation and a stark shortcoming in (re)playability. "Playability is a term used in the design and analysis of video games that describes the quality of a video game in terms of its rules, mechanics, goals and design. It refers to all the experiences that a player may feel when interacting  with  a  game  system." \cite{sanchez_playability_2012} We have appended "(re)" to the playability term as our goal was to not only encourage initial player engagement, but we also wanted to explore the player experience upon repeat plays of the games for our case studies. This dichotomy lead to an observation that games for education requires a balance between simulation and gameplay. Discussion of this observation provokes a redevelopment of the gaming experience which pulls from the game design researcher’s knowledge of engaging the player and the bioinformatics researcher’s knowledge of the medical field.

\section{Background}
VR is a developing field of technology which portrays  a  virtual  space or collection of objects as an interactive world \cite{sherman_chapter_2019}. The technology  immerses  users  in  a virtual environment  and provides sensory  feedback  to invoke a feeling that the created environment is a real physical environment. The sensory feedback provided by VR empowers learners to be an active participant in their education, making the tool ideal for scholarship \cite{bowman_educational_1999, pantelidis_reasons_2010}. VR has been applied to many different educational settings including pilot training \cite{yavrucuk_low_2011}, urban planning \cite{bowman_educational_1999}, physics education \cite{loftin_applying_1993} and primary school math and science education \cite{johnson_augmenting_2002}.

\subsection{VR in a Clinical Setting}

Recently, VR instruction expanded beyond the conventional curriculum to the clinical setting for use by both patients and physicians \cite{huang_investigating_2010, jimenez_patient_2018, pandrangi_application_2019, kuehn_virtual_2018, reznek_virtual_2002}. In 2000, VR made its debut in health care when SnowWorldVR was developed to act as an alternative analgesic for burn patients \cite{hoffman_use_2000}. This original application of VR to medicine paved the way for several more applications developed for pain treatment, physician education and patient education \cite{kuehn_virtual_2018, reznek_virtual_2002, bridge_development_2007, birnie_usability_2018}. Increased implementation of VR in a clinical setting has been shown to decrease patient anxiety by increasing patient understanding and by acting as a distraction during painful procedures \cite{jimenez_patient_2018, hoffman_use_2000, birnie_usability_2018, hoffman_virtual_2011}. While the technology was originally applied to areas of general medicine, recent applications in pediatrics have shown similar effects, decreasing patient anxiety and pain \cite{birnie_usability_2018, atzori_virtual_2018}. 

Unfortunately, most VR applications developed for medicine focus fully on the gaming experience \cite{bridge_development_2007}, giving them little educational value, or are replications of medical procedures \cite{jimenez_patient_2018, kuehn_virtual_2018, reznek_virtual_2002} making them boring for children . To our knowledge, there has been no VR experience developed to educate young patients about their illness by combining both simulation based education and the domain of Games User Research (GUR).

\subsection{Games User Research}

Games User Research (GUR) is a branch of the broader area of Games Research. GUR focuses on the use of research methods drawn from Human Computer Interaction (HCI) research, while integrated with the games specific parameters. For example, in HCI user experience studies do not rely as heavily on engagement or entertainment factors as much as GUR do. Drachen, Mizra-Babaei, and Nacke, in their opening chapter to “Games User Research,” define GUR as “an interdisciplinary field of practice and research concerned with ensuring the optimal quality of usability and user experience (UX) in video games.” \cite{drachen_gur_2018}

 The challenge with GUR is that “[if] game development were an ancient temple, the three biggest central pillars would be design, art, and programming.” \cite{drachen_gur_2018} Contributing to these central columns would be almost every other discipline including sociology, psychology, biology, medicine, earth sciences, law studies, economics, and almost every other form of research. GUR is like “a strong vibe that has its tendrils spread across the vast majority of columns, supporting each of them at the same time…” \cite{drachen_gur_2018} Because of the breadth of analysis that is considered in GUR, there are a growing number of research methods that researchers will use to devise an accurate and appropriate test of game playability. In Chapter 7, “An Overview of GUR Methods,” Medlock discusses two approaches to curating specific GUR playability tests: cookbook and structural. \cite{drachen_gur_2018} For our purposes, we realized that because we were working with a completed version of the game, we needed to consider post-launch GUR techniques. In his chapter on post-launch GUR, Ian Livingston, introduces the use of Grounded Theory as a method for researching after the project has wrapped up \cite{drachen_gur_2018}. "Grounded Theory is as systematic qualitative methodology where a theoretical framework is developed through the analysis of a rich qualitative data source." \cite{livingston_post-launch_2018}

In our case, we had a version of our game, LeukemiaSIM, that was ready for revision. The game, in its initial form, aimed to educate pediatric leukemia patients and their families about leukemia physiology and pathophysiology. The primary version of LeukemiaSIM consisted of a school environment where players could interact with nine unique blood cell characters and play mini games that reflected the physiology of each cell. There was also a section where players could experience unhealthy blood cell division under cancer pathophysiology and play mini games to "treat" the cancer. For our revised edition of the game, we focused on the main gaming environment where players interacted with the blood cell characters and the red blood cell oxygenation mini game, because we felt these two scenes were applicable to the broadest audience and would have the largest impact on patient understanding of leukemia physiology. Focusing on these game scenes also gave us a chance to take a reflective position on the research, while at the same time, using these reflections to design and rationalize new creative directions for the use case games we examined without having to examine the whole game.

\section{Process}
\subsection{Overview}
In this paper we explore two case studies that are part of a larger set of games and interactive experiences to help children better understand Leukemia called LeukemiaSIM. The two case studies are mini-games that appear in the context of LeukemiaSIM. This first, the Red Blood Cell Oxygenation Game, was created to help the player better understand how blood cells oxygenate as they move from the lungs to the heart before their journey on to the rest of the body. The second, the Blood Cell Interaction Zone, was created to explore the cellular makeup of human blood by anthropomorphising the various types of cells in context with the role they play. Each of these mini-games targeted specific knowledge based learning outcomes identified by the research team. With both case studies, we review the current state of the games, ask questions about the efficacy of the interactions and gameplay, and devise improvements that would serve to enhance gameplay engagement, encourage repeat playability, and continue to align with the learning outcomes. 

\subsection{Case Study 1: Red Blood Cell Oxygenation Game}
The Red Blood Cell Oxygenation Game was one of the first mini games designed for LeukemiaSIM. The goal of the game was to collect oxygen as you travelled through a blood vessel. In designing the game, we wanted users to understand the role of red blood cells in oxygen transport and looked to provide a visual metaphor for the journey taken by the red blood cell as it travels from one end of the heart to another. The original version of our Red Blood Cell Oxygenation Game consisted of travelling through a continuous vessel in a magic school bus as an homage to the famous series of books and animations called "The Magic School Bus". \cite{cole_magic_2020} The player's goal was to collect as many characters representing oxygen as possible before crashing into three obstacles that appeared at set intervals. 

Starting point - since the game was already a well-planned, and implemented experience, our first session was an overview of the mini game and an exploration of what was working and what was not from the perspective of someone who had never played the game before. We identified a few areas that were unclear or not effective. We also asked the question, when is a visual metaphor appropriate and when is it better to use the real object (i.e. the red blood cell; oxygen molecules)?

In game design and production, creative ideas reveal themselves or die off throughout the design and development process. Authors refer to the “iterative design” process as a general knowledge concept and will typically outline this process in similar ways. \cite{fullerton_game_2013, flanagan_critical_2009} This process follows the basic cycle: come up with an idea, build it, playtest it, reject or keep it, repeat. 

We followed this iterative process and integrated some analysis to idea conception. During the first meeting, the game design researcher, suggested “returning back to basics” and began asking some questions about the game. These questions aligned to analyses derived from the “10 Usability Heuristics Applied to Video Games” from the Nielsen Norman Group \cite{joyce_10_2019}. Specifically, we addressed the following outcomes:
\begin{enumerate}
    \item What is the demographic of the player?
    \item What is the purpose of the game? 
    \item Why would the player want to play the game? 
    \item Why would the player want to play the game more than once?
\end{enumerate}

\begin{figure}[!ht]
    \begin{center}
  \includegraphics[width=0.75\textwidth]{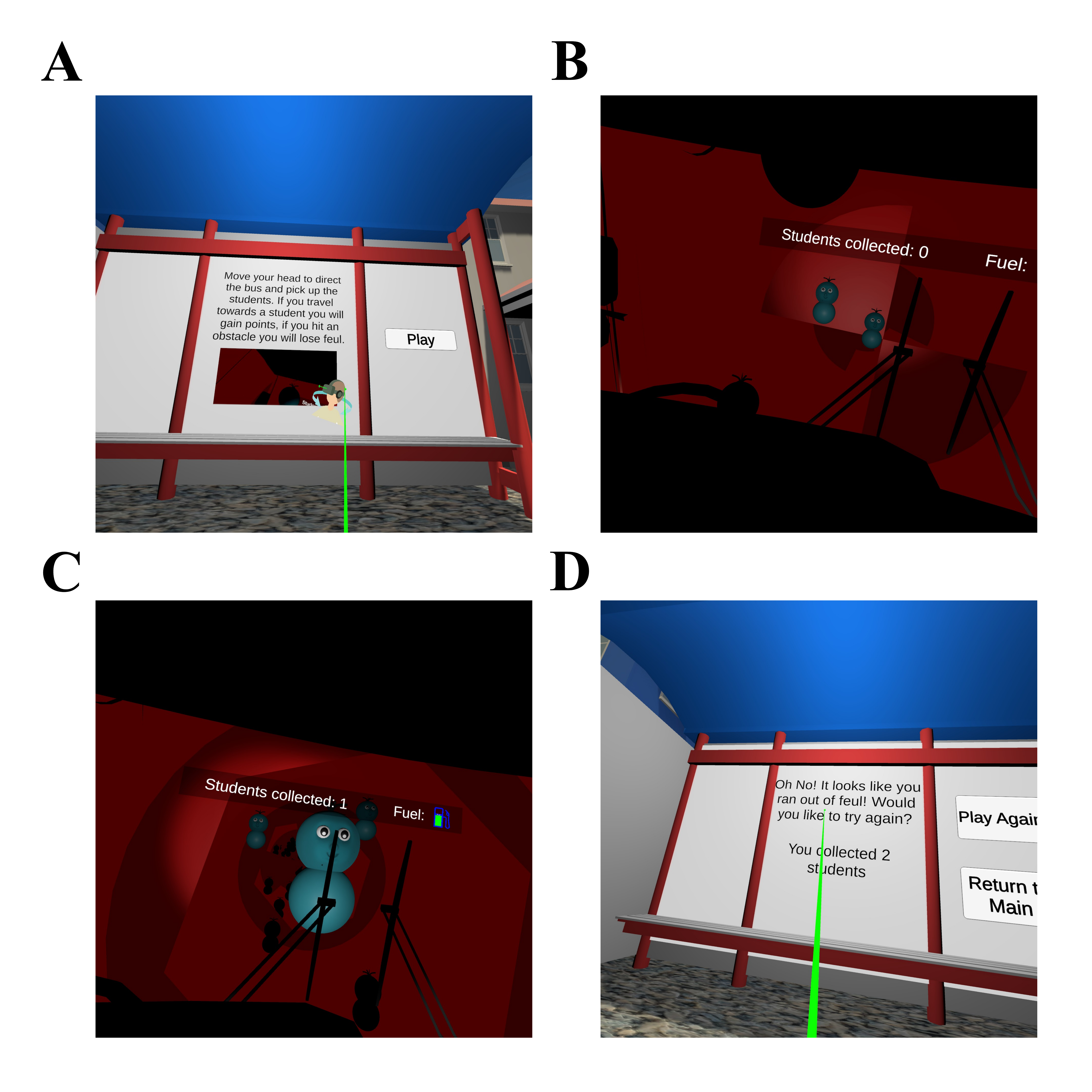}
  \end{center}
  \caption{The original Red Blood Cell Oxygenation Game. a) The menu and instruction screen seen by the player when they first enter the game. b) A visualization of the blood vessel transport. c) Illustration of a player about to collect an oxygenation character. d) The finishing screen viewed by the player once they reached the end of the level.}
  \label{fig:OrignalRBC}
\end{figure}

Thoughts about these questions started a conversation that identified inconsistencies between the intention of the game and the realized gameplay experience. We discovered the following challenges within the first implementation \textbf{(Figure \ref{fig:OrignalRBC})}:
\begin{itemize}
    \item The concept of driving a bus in a blood cell to gather oxygen molecules was a big narrative stretch that required significant backstory and exposition. 
    \item The gameplay was not very exciting. The sensation of movement was lacking in the gameworld and the physics of driving a huge bus dragged on the whole experience.
    \item The learning outcomes were external to the gameplay rather than embedded in the interactions in the game.
\end{itemize}

We began to ask some speculative questions about how we could better address the players’ sense of fun and engagement. Additionally, we wondered how we could find authentic ways of embedding the learning outcomes into the gameplay. We started asking:
\begin{itemize}
    \item What if we did not use a bus and instead the player was controlling a red blood cell to pick up the oxygen? What would that feel like? 
    \item What if we could make the movement through the blood vessel feel faster? 
    \item What sort of gameplay mechanics are going to be the most interesting? A continuous play with items gradually getting harder to gather, or a level by level with a beginning and ending?
    \item What if the goal was to oxygenate as many red blood cells as possible as you pass by the lungs on your way back to the heart and into the rest of the body?
\end{itemize}

We also discussed a change to the playable dynamic. We noted that the idea of driving a bus down a blood vessel and running over oxygen characters to pick them up seemed to have some potential unintended consequences \textbf{(Figure \ref{fig:OrignalRBC}b)}. A bus is big and cumbersome. The act of driving over characters seems cruel. So, we decided to brainstorm some ideas about what we could do to make the gameplay more exciting. The game designer, suggested a skating or surfing metaphor. Being in the blood flow, and guiding yourself to the oxygen seemed more action oriented. This resonated for the bioinformatics researcher, who mentioned that surfing was one of the original ideas she had. Then we happened upon the idea that the player could be surfing on a red blood cell. This meant we could also create a simulation of the actual interaction between the oxygen and red blood cells binding. Now the game became a challenge to oxygenate as many blood cells as possible. Each cell could only trap four oxygen molecules, so the player would jump from one cell to another and rack up as many blood cells as possible in the trip from the lungs to the heart. 

\begin{figure}[!ht]
\begin{center}
  \includegraphics[width=\textwidth]{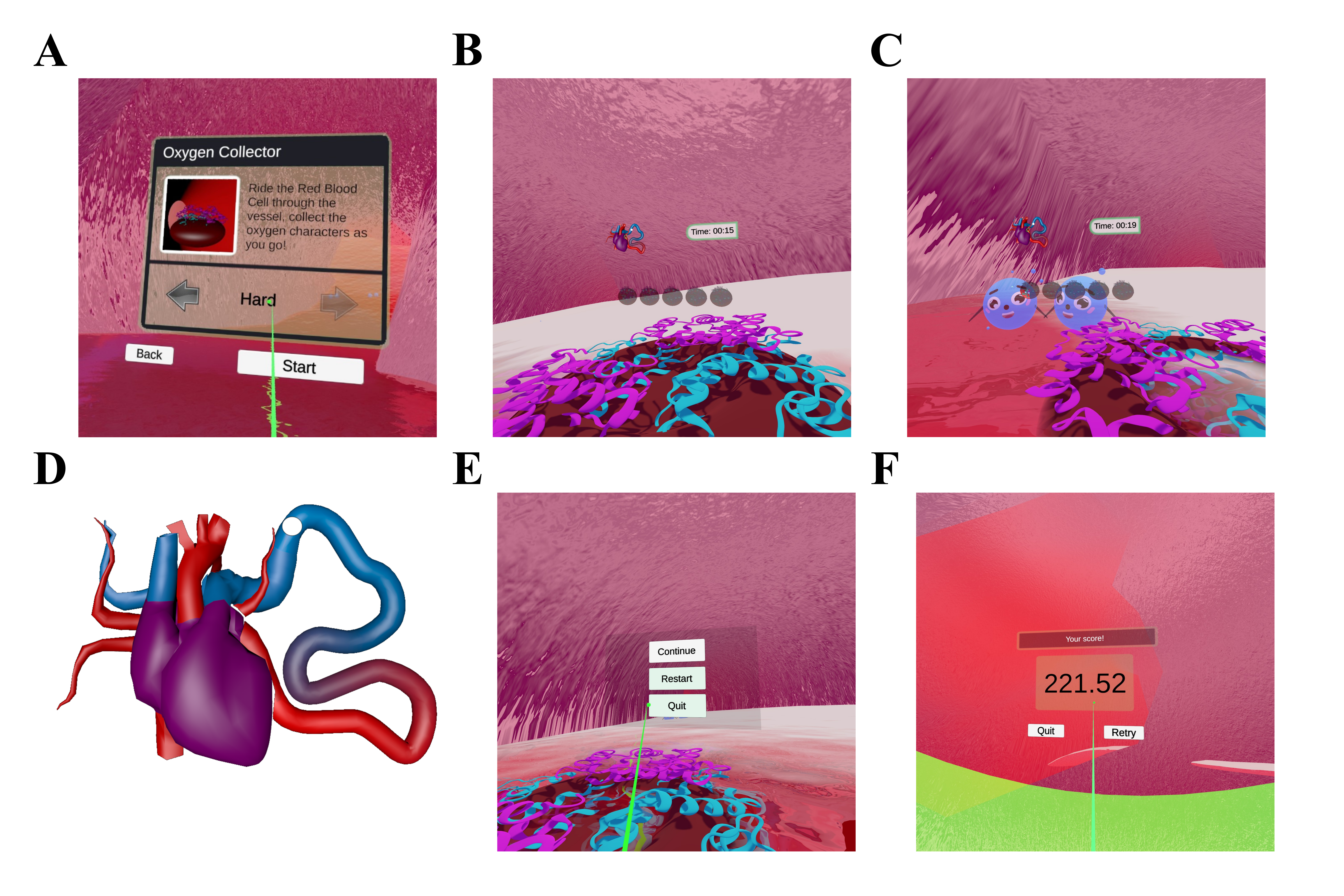}
  \end{center}
  \caption{The updated Red Blood Cell Oxygenation Game. a) The menu and instruction screen seen by the player when they first enter the game. b) A visualization of the player riding the blood cell through the vessel. c) Illustration of a player about to collect an oxygenation character. d) The mini map illustrating where the player is in the vessel. e) The pause screen which allows a player to return to the main area. f) The finishing screen viewed by the player once they reached the end of the level. Notice the player score is displayed.}
  \label{fig:UpdatedRBC}
\end{figure}

Our discussion resulted in a new set of gameplay rules, environment design choices, and character elements. First, we modified the surroundings of the player to better reflect veritable red blood cell transport during oxygen collection. In the original iteration of the game, the player was carried through a continuous forward facing tube that was generated at runtime \textbf{(Figure \ref{fig:OrignalRBC}b,c)}. This surrounding was not representative of red blood cell locomotion because blood vessels grow, shrink, wind and diverge [25]. We redesigned the play course to reflect a true blood vessel \textbf{(Figure \ref{fig:UpdatedRBC}b)} by modifying a model of the human heart. In particular, we attached a mesh to the human heart model, connecting the left pulmonary artery to the left pulmonary vein with a hollow surface the player could ride through. This new "blood vessel" followed a winding pathway that was positioned around the heart model where the lungs would be located. The vessel ended in a natural and physiologically relevant final destination: the left ventricle of the heart which acts as a pump passing oxygenated blood to the aorta \textbf{(Figure \ref{fig:UpdatedRBC}e)}. The twists and turns of this new vessel were not only more engaging for the player, but the final destination gave a natural new objective: to collect all the characters representing oxygen before reaching the final destination: the left ventricle. This new objective allowed us to remove the obstacles, further improving the anatomical correctness of the pathway.

We then changed the method of player transport to a red blood cell that the player could surf on as they went through the vessel. We designed our red blood cell to have both a cartoon surface structure, shown in most red blood cell representations, and a hemoglobin protein structure with four heme groups to attach oxygen to taken from an x-ray crystallography study[26]  \textbf{(Figure \ref{fig:UpdatedRBC}b)}. This new outlet for grounding the player enhanced the learning outcomes of the game by showing scientific representations of the red blood cell and by allowing for correct placement of oxygen characters on the hemoglobin structure during gameplay. Additionally, the movement of the player on top of the red blood cell gave a new feeling of motion that was more exciting and less claustrophobic then the original bus movement.  

Next, we worked to improve the environment of the Red Blood Cell Oxygenation Game by modifying game materials and textures \textbf{(Figure \ref{fig:UpdatedRBC})}. We began by changing the base of the vessel to a water texture coloured red. This water texture refracted light from the skybox giving the appearance of a translucent and pliable duct rather than a sedentary tube. We added a layer of flowing red water to the bottom of the vessel to simulate natural liquid motion. We tried simulations in which the player could dive under this water layer, however, the playability of the game was enhanced by having the blood cell remain on top of the water. This grounded the player in a known environment and  a feeling of movement, invoked by the water texture, was sustained (this movement feeling was not as strong when the player went below the surface of the water). We deliberately decided that playability was more important than the physiological correctness of being in the blood liquid layer, and removed any abilities to go under the water. The texture of the skybox was also modified to look like blood vessels to further exemplify the feeling of being in a human body. 

We also modified the text interface, which outlined game rules and served as an access point for exiting the Red Blood Cell Oxygenation Game, to make it more user friendly and straightforward. We reduced the original instruction text \textbf{(Figure \ref{fig:OrignalRBC}a)}, which was long and confusing, to a single line \textbf{(Figure \ref{fig:UpdatedRBC}a)}. Buttons were moved to the bottom of text interfaces to reduce confusion and allow for a natural reading flow. Lastly, all interactive screens were simplified by reducing or removing images and limiting the number of elements that could be interacted with \textbf{(Figure \ref{fig:UpdatedRBC}a,e,f)}. 

\subsection{Round Two! Further Iterations}

Once, these changes to the Red Blood Cell Oxygenation Game were implemented, the bioinformatics researcher and the game designer met again to discuss how the game could be further refined. Our second round of discussions began by asking open ended questions focused on describing the demographic and objectives of the game. This is a variation of Game User Testing Heuristics [28] which was modified to fit our conversation needs. The game designer asked the bioinformatics researcher to take the whole thing back to the basics of why she built the game in the first place. Our discussion was not about justifying the choices made in the current version of the game, but more about reflecting on the original motivations. We discussed who the game was made for?  What the game was about and what we wanted the player to remember about the game? Why a player would want to play this game? And, where the game would be played? At home? At school? At the hospital?

The bioinformatics researcher explained that the entire LeukemiaSIM experience was developed for children aged 9 - 12 currently battling or in remission for acute lymphoblastic leukemia, and the Red Blood Cell Oxygenation Game would also fit this demographic. She described that the secondary audience for the game were interested family members who wanted to become educated about acute lymphoblastic leukemia. The bioinformatics researcher saw the video game being played primarily with the assistance of a child life specialist describing the learning outcomes according to their professional training. Consequently, the game was to be generally distributed to and played in hospitals, but we also saw potential for patients to play the Red Blood Cell Oxygenation Game in their own homes. The primary goal of the game was education on red blood cell oxygen transport and anatomy. Literature has shown that enjoyment of learning leads to greater retention, so we also aimed to make an enjoyable but challenging experience \cite{giannakos_enjoy_2013}. We concluded that players would most likely be playing the game to distract themselves from chemotherapy treatments and the game needed to be sculpted accordingly. 

Our discussion led to thoughts about the types of games and the challenges that players expect to deal with. For example, a game that has a beginning, middle, and end, like a Super Mario Bros. level, will have different design needs over an infinitely increasing difficulty game like Flappy Bird or Tetris, where you play until you gain the highest score or die. We decided that a game with a beginning, middle, and end was most appropriate for our experience because red blood cells have a limited amount of time to oxygenate as they pass through pulmonary arteries en route to the heart and opted for a timed game to replicate this scenario. 

To add challenges to our timed game we implemented a scoring mechanism based on the number of oxygen characters collected en route to the heart and the amount of time it took the player to reach the heart \textbf{(Figure \ref{fig:UpdatedRBC}f)}. We gave players visual cues of their score along the way through a display with a clock and red blood cells that filled out as characters were collected \textbf{(Figure \ref{fig:UpdatedRBC}b)}. At the end of the game, we modified the closing screen to display a player's score, ultimately increasing replayability by giving a challenge for the next time around. Replayability was also increased by incorporating two levels of difficulty which varied baseline speed and oxygen placement, allowing for players to advance from one level to the next as their skills improve. 

In this new iteration of gameplay we further incorporated a minimap \textbf{(Figure \ref{fig:UpdatedRBC}d)} to ground players in space and to show an exterior view of the vessels. We designed the mini map to mirror anatomical drawings of the heart, further increasing the educational value of the game. On the mini map the player was represented with a white dot on a flat figure showing the circulation. This figure transitioned from blue to red as oxygen was collected, just as anatomy figures would be drawn in a textbook.

\begin{figure}[h!]
\begin{center}
      \includegraphics[width=\textwidth]{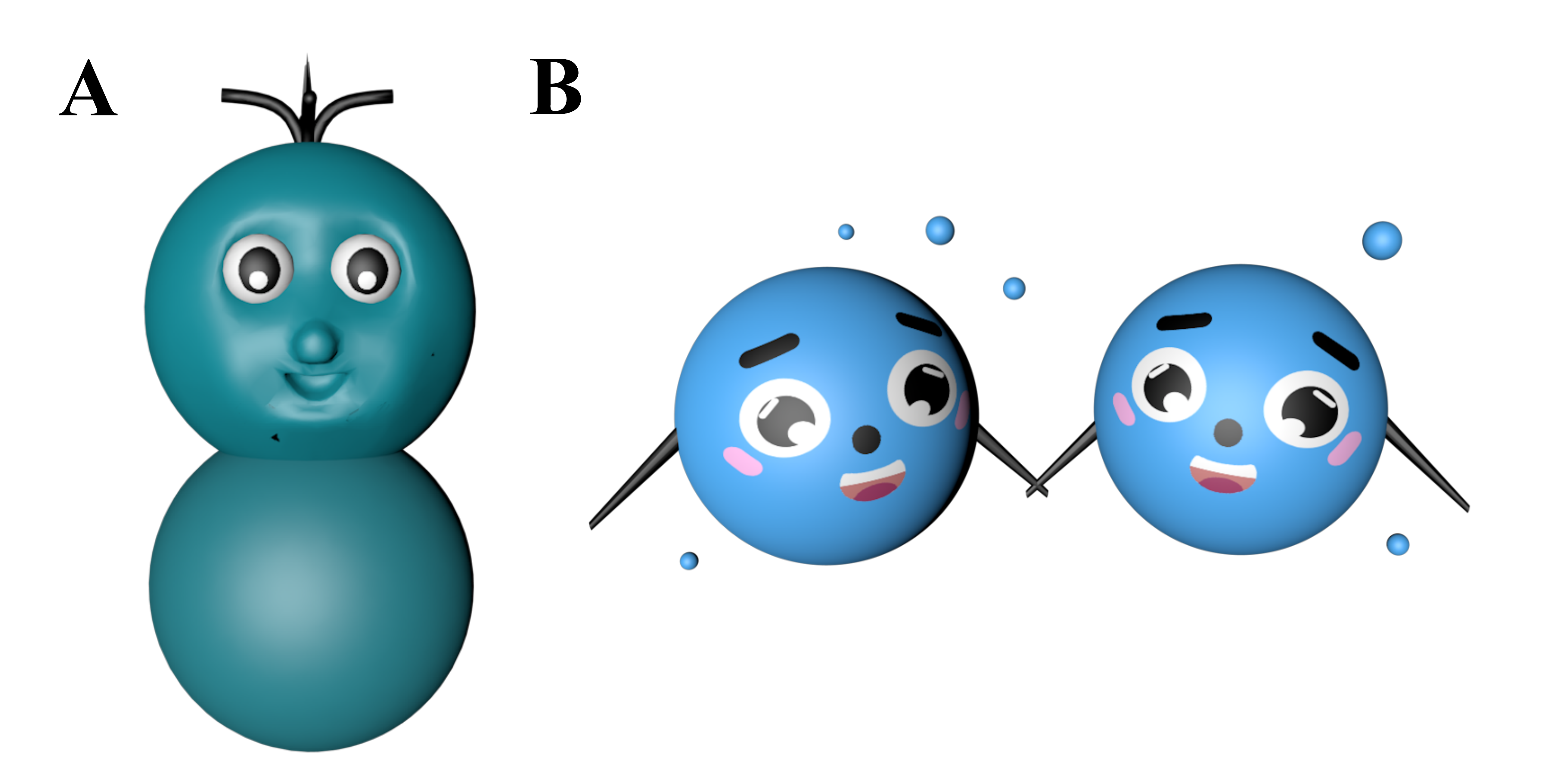}
\end{center}

  \caption{The oxygen character refinement. a) The oxygen character in its original form which we noticed looked a lot like a snowman. b) the updated oxygen character, refined to remove ambiguity.}
  \label{fig:Oxygen}
\end{figure}

With this new and improved gameplay, our discussions began to naturally move from the major gameplay elements, to fine-tuning. We noticed that the figures meant to represent oxygen looked too much like snowmen \textbf{(Figure \ref{fig:Oxygen}a)}. Out of concern for ambiguity and fear that children might think that red blood cells deliver snow rather than oxygen, we redesigned the characters to have a side-by-side orientation in which two separate figures are holding hands \textbf{(Figure \ref{fig:Oxygen}b)}. We also tweaked some of the textures to further exemplify the feeling of movement through a vessel. 

As further development opportunities present themselves, we can see a number of production quality enhancements that will improve the experience. It is important to note that without correcting the major game design problems, refining artwork or enhancing frame rates would not have made the game more engaging and playable.

Part of having an effective iterative cycle of production is knowing what features need attention in what order of priority. It’s a bit like carpentry. Putting in the cabinetry does not make much sense if the framing of the walls is not complete. Knowing what fixes should take precedence over others is key to a successful iterative production process.

With the crucial features implemented and a testable version of this game available, we moved on to the next case study in order to take our fresh experience in our iterative production process and apply it to another interactive experience: the Blood Cell Interaction Zone.

\subsection{Case Study 2: Blood Cell Interaction Zone}

\begin{figure}[!ht]
\begin{center}
      \includegraphics[width=0.75\textwidth]{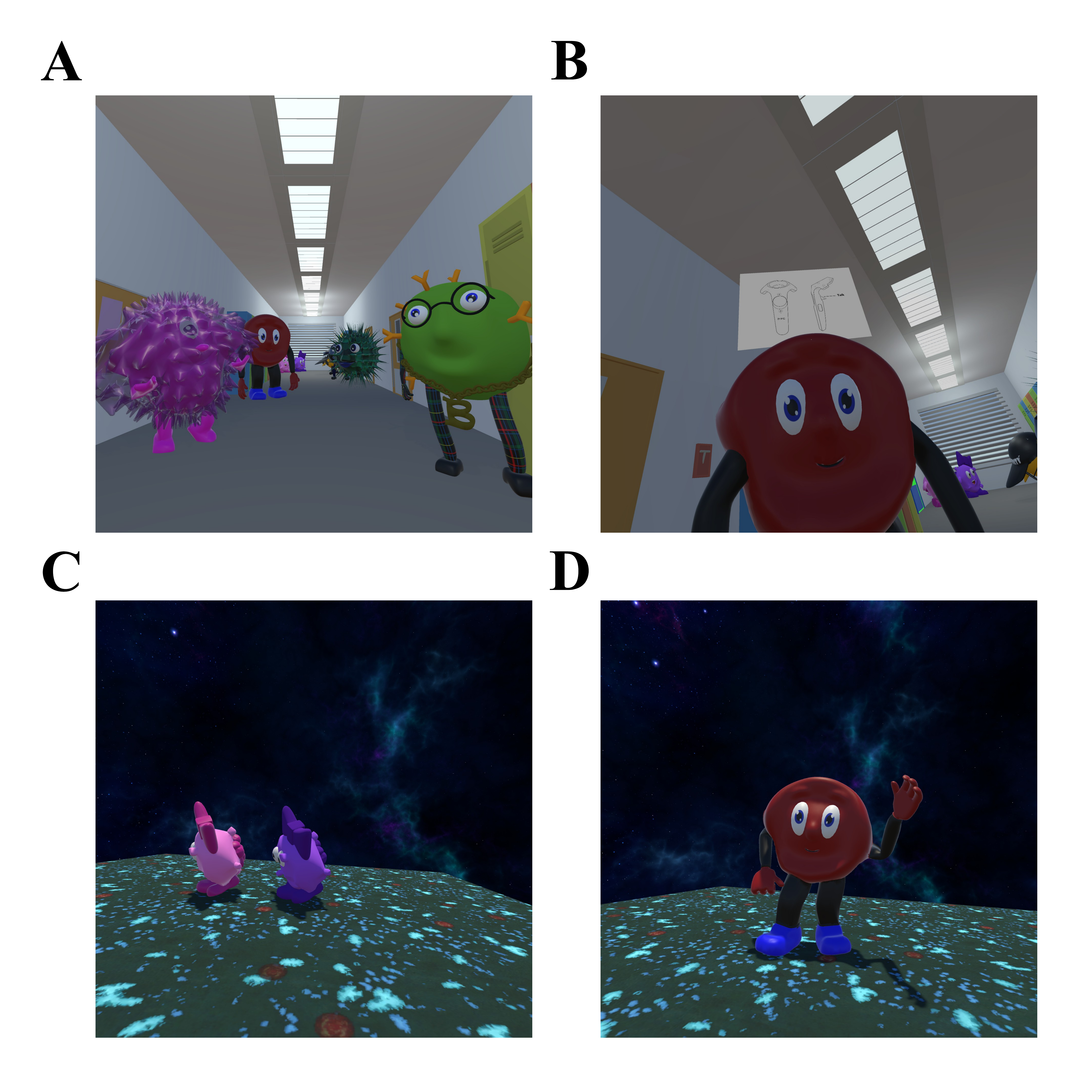}
\end{center}

  \caption{The original Blood Cell Interaction Zone. a) The main interaction hallway with the 7 blood cell characters. b) Interaction with the red blood cell character. Shows interaction to interact with the red blood cell character. c) Interaction with the megakaryocyte cell characters. d) Interaction with the red blood cell character.}
  \label{fig:Hallway}
\end{figure}

To introduce players to the different kinds of blood cells and their role in leukemia, an area was set up within LeukemiaSIM where players could interact with seven important blood cells \textbf{(Figure \ref{fig:Hallway})}. Originally, this environment was designed as a school hallway where players could swing their remotes in a running action and walk up to characters designed to represent the blood cells \textbf{(Figure \ref{fig:Hallway}a)}. Once facing the characters, players could press a button on their remote and listen to the characters speak metaphorically about their role in physiological processes including oxygen transport, clotting and host-immunity \textbf{(Figure \ref{fig:Hallway}b,c,d)}.

We decided to use this zone as our second case study because the environment lent itself well to more conventional forms of instruction, which Wickens suggests are essential to learner retention of experiential learning in VR environments [27]. Since changes to all levels could not be made at this stage, by ameliorating this environment, we felt we would have the largest impact on player education. Additionally, the area acted as a home base where players could access other mini games, including the aforementioned Red Blood Cell Oxygenation Game.

We followed a similar iterative process to the one outlined in Case Study 1 when refining this space. We started by asking questions to answer who, what, why and where the interactive was going to be played. The game designer again interviewed the bioinformatics researchers to derive the purpose of the interaction and what we were looking at. Questions included who would be interacting with this environment? Where would the player be situated when interacting with the characters? Why should the player have this interaction? And, when would the interaction take place?

The bioinformatics researcher explained that this interactive was developed to initiate player recognition of various blood cells present in a drop of blood. We noticed the school hallway was a large narrative step as the environment was not representative of blood. The game design researcher mused over the idea of making the surrounding more realistic and actually moving through a blood drop. We discussed this idea, but noticed that gameplay navigation might be a challenge because a natural drop of blood would have cells above the player which could be hard to reach. Then the game design researcher suggested putting the blood sample on a slide which the player could interact with in two dimensions. 

This led to a conversation about changing the environment to a lab. We noticed that children have a lot of fun of interacting with lab equipment such as microscopes and knew this idea would be more visually interesting than using an already too familiar classroom setting. We discussed the gameplay of the environment and came up with a new scenario where the player would interact with a microscope with zoom buttons. Once the player zoomed in several times they would enter the microscope and begin to interact with the blood cell characters. We ended by pondering the educational component of the interactive and the need to provide a way to get to the mini-games. In order to make this a more playable level, we chose a “collect them all” approach to finding and identifying all of the types of cells in the sample on the slide. 

\begin{figure}
\begin{center}
      \includegraphics[width = \textwidth]{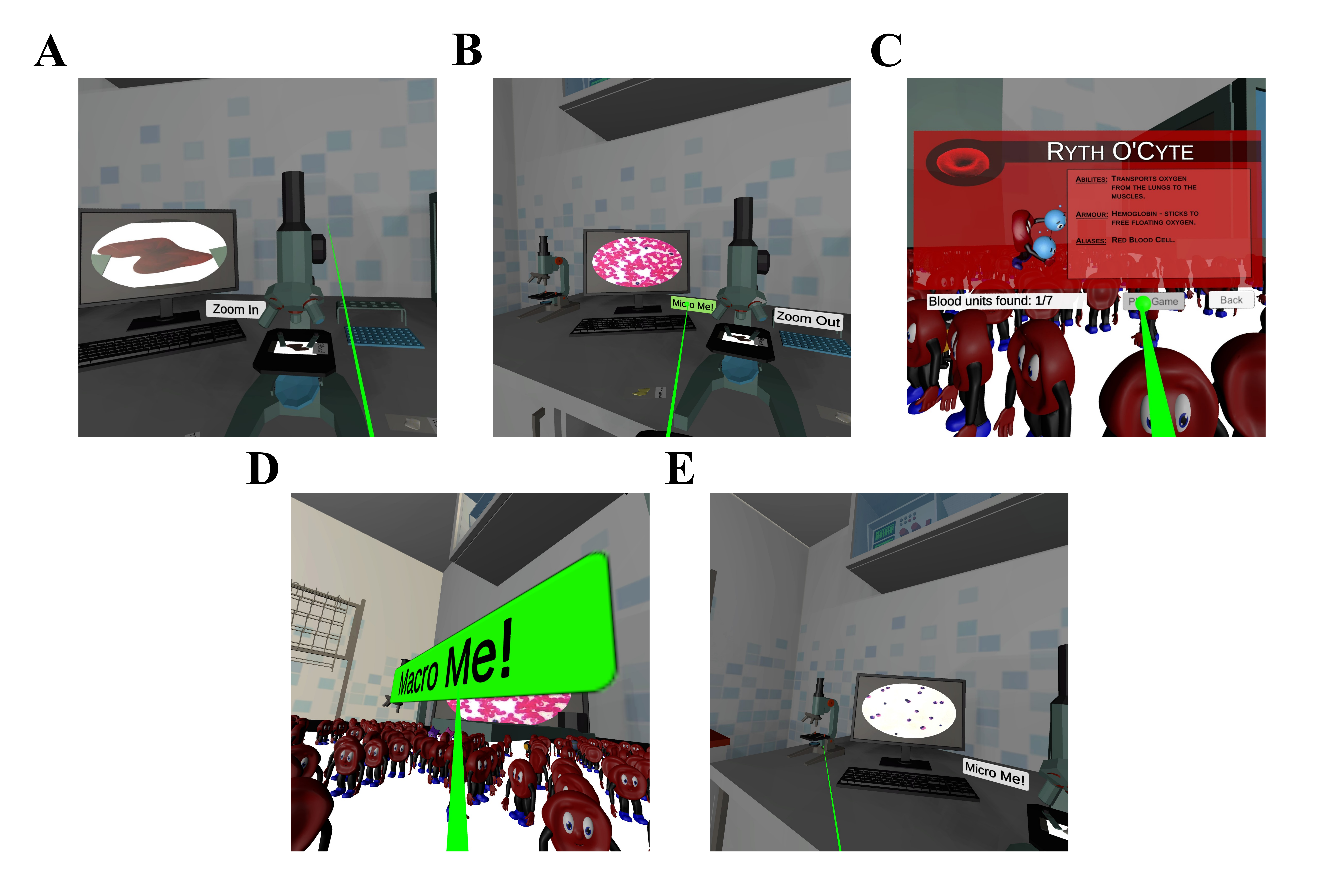}
\end{center}

  \caption{The updated Blood Cell Interaction Zone. a) Interaction with the microscope object in the laboratory area. b) Interaction with the zoom buttons of the microscope interaction. Computer screen shows zoomed in cells. c) Red Blood Cell cell character description and interaction. d) Interaction with the "macro me" button and display of the character in micro form. e) Computer screen displaying the plasma slide.}
  \label{fig:lab}
\end{figure}

We began our modifications of this game with significant changes to the setting \textbf{(Figure \ref{fig:lab})}. We found a model of a typical biology lab online and adjusted the surroundings for our purposes. We isolated a microscope in the environment and attached an interactive computer screen which showed blood cells as various zoom distances \textbf{(Figure \ref{fig:lab}a,b)}. We created a model of a microscope slide with a blood droplet and placed it under the microscope. We mirrored this blood droplet onto the computer screen and added zoom in and zoom out buttons. When the zoom in button was pressed, we changed the image on the computer screen to a veritable blood microscopy sample. We added the capacity to zoom into the blood a second time, to further amplify the images. Once the images were at full amplification, we added functionality for the player to enter the blood cell slide \textbf{(Figure \ref{fig:lab}b,d)}.

Once the player was inside the blood cell slide, we instantiated a new interactive with the blood cell characters. In this zone, a green highlight appeared on each of the characters when the player pointed towards them. The player could then click on these characters to bring up the virtual information card \textbf{(Figure \ref{fig:lab}c)}. On these cards we scribed the roles of each blood cell using descriptions analogous to those on a playing card. Items included character names, abilities, armour and aliases. Each of these descriptions mirrored true blood cell physiology. For example, the red blood cell card had the name Ryth O'Cyte to represent erythrocyte (the scientific name for a red blood cell). The abilities on the card were "Transports oxygen from the lungs to the muscles", the armour was hemoglobin and the alias was "Red Blood Cell". We also added a button to the bottom of these interactive surfaces giving the ability to enter the mini game of the character.

Once the lab environment had been created and interactions with the blood cell characters were established, we met again to discuss how we could refine the environment. In this meeting we focused on the playability of the environment, asking questions such as will interacting with a computer screen make the microscope analogy confusing? How will players know how many blood cells there are to collect? Is there a way to implement levels into our interaction for younger players? What does locomotion within the gaming environment look like? 

These questions led to a discussion about refining elements of the play environment. We discussed moving the zoom in and zoom out buttons from the computer screen to the microscope to illustrate that the player was interacting with the miscroscope not the computer. The game design researcher, also suggested we might have multiple microscope slides for different levels of play. This gave the bioinformatics researcher the idea of creating slides for the different blood components. Often when researchers analyze blood samples they will centrifuge the samples to separate white blood cells from red blood cells and plasma. The bioinformatics researcher saw this process as an ideal way to separate out the less common white blood cells from the abundant red blood cells in a manner that was scientifically accurate. Younger players could then select the white blood cell slides to find these less common cell types if they were having a hard time accessing the various characters in their initial experimentation with the common blood droplet.

We also discussed how integration of locomotion into the gaming environment could help players find the blood cell characters. The bioinformatics researcher was concerned with integrating locomotion because her experience with children in VR environments suggested they did not understand typical VR movement actions. This led to a conversation about other forms of locomotion and scientifically appropriate forms of locomotion in our micro-sized simulated environment. The game designer had the idea of adding random player movements, much like random molecular translational motion might occur on a blood slide. 

These discussions led to a second iteration of the Blood Cell Interaction Zone. First, we moved the interactive zoom in and zoom out buttons next to the microscope to illustrate that the player was interacting with the microscope not a computer \textbf{(Figure \ref{fig:lab}b)}. This re-positioning of the buttons, not only increased the playability of the environment but also led to a happy accident. When we moved the buttons we kept them the same size in the blood cell interactive as when the player was in the microscope talking to the blood cell characters. This change in scale led to giant buttons that further conveyed the experience of being tiny \textbf{(Figure \ref{fig:lab}d)}. The large microscope buttons (now used to exit the micro environment), coupled with the giant laboratory equipment, exemplified the pretense that the only thing that changed size was the player. 

In this iteration, we added slow random movement so that the player may see more of the environment when hunting for characters. This gave the player multiple perspectives and increased the likelihood that the player would come in contact with less common blood cell characters. We added a count to the bottom of the interaction screens to make the player aware of the characters they had already found and how many more they needed to discover. Finally, we created four different blood samples: pure blood, red blood cells, white blood cells and plasma, which could be exchanged to change the number and types of blood cells in the environment \textbf{(Figure \ref{fig:lab}e)}.

Once these new changes were implemented, we met a final time to refine the gaming experience. In this final meeting we talked over the artistic elements of the environment that would enhance the interactivity of the zone. Suggestions included incorporating animations that mirrored the actions of the different blood cells into the environment and the addition of voice commands to enhance instruction.

While our newest version of the gaming experience does not yet include voice commands, we did create animations to reflect the abilities of each blood cell character. For example, the red blood cell character card now has an illustration of the red blood cell character carrying the oxygen characters developed for the Red Blood Cell Oxygenation game. 

Like in our refinement of the Red Blood Cell Oxygenation game, we did not consider refining artwork and other production quality elements at this stage. We were focused on correcting major design elements, and knew that once these elements were fixed, production quality could easily be added to the gaming experience.

\section{Results}

Code for the original and updated versions of these two LeukemiaSIM case studies can be accessed from GitHub: \url{https://github.com/ebrintn/LeukemiaSIM}. To modify the programs Unity version 2019.1.12 (\url{https://unity3d.com/get-unity/download}) and Steam VR (\url{https://store.steampowered.com/app/250820/SteamVR/}) are required. Users will also require VR hardware including a Vive Pro headset and a computer with a graphics card with DX10 capabilities. Once software and hardware requirements are met, players can run and modify the systems within the Unity development kit by selecting either the "OriginalSimulation" or "UpdatedSimulation" folder in the game selection window of Unity. Pre-compiled  original and updated games can respectfully be accessed from \url{https://ebrintn.itch.io/leukemiasim-originalversion} and \url{https://ebrintn.itch.io/leukemiasim-updatedversion}. These compiled versions of the games are currently only for the a Windows PC environment.

When comparing the two gaming experiences players and developers will notice several key elements that increased the playability and educational value of our VR experiences. Players should notice a refinement of scientific concepts from the original version of the LeukemiaSIM simulations to the updated version. Originally, when we were developing LeukemiaSIM we used a lot of analogies that we thought would make the game more approachable for our young audience. In the redesign process, we noticed that many of these analogies made the games confusing and drew attention away from the main learning objectives we were trying to incorporate into the games. By returning to verified science, our modifications not only made the games more applicable to future scientific inquiry but removed any ambiguity invoked by misinterpretation of analogies. This was an issue originally brought to our attention by Child Life Specialists who work directly with patients to help them understand their illnesses.

Similarly, the next major change that players will notice is the use of more authentic models in the updated version of the LeukemiaSIM games. In our original version of LeukemiaSIM, we developed a gaming experience that mostly drew from our own life-experiences, a phenomenon characterized by media theorist Marshall McLuhan as "rear-view mirrorisms" \cite{mcluhan_war_1968}. It is the notion that we naturally gravitate to old ways of doing things when we encounter new technologies. In our own life we had been to school and on a bus, and thought to incorporate these natural environments into this new virtual surrounding. However, in VR there is a suspension of disbelief that allows us to enter environments that we have yet to experience in reality. We are willing to suspend our disbelief because the VR environment is inherently a magical space where normal rules don't have to apply. In the new version of LeukemiaSIM we drew on this suspension of disbelief to put the player inside a microscope and on a blood cell. In the new environment, players could learn from their surroundings in addition to learning through interactions, increasing the playability and educational outcomes of the game.

Finally, in redesigning the Blood Cell Oxygenation Game and the Blood Cell Interaction Zone we looked to implement games that could be played multiple times. In the Blood Cell Oxygenation Game the replayability manifested itself as a challenge to beat one's high score. In the Blood Cell Interaction Zone, we added a gaming element of trying to find all the blood cells and created multiple blood slides that could be exchanged to make finding the characters harder or easier.By making our environment and the mini-games replayable, learning outcomes could be reinforced with better knowledge retention. 

As a primary playtest for the updated LeukemiaSIM environment, one of our researchers who had seen the LeukemiaSIM game but had never been in the experience entered the game. This researcher suggested that the modifications to the game did make the experience more accessible and interesting. In particular, he really enjoyed the change from the bus to the blood cell and was pleasantly surprised when he was put inside the laboratory environment. However, it was suggested that gaming objectives were still ambiguous and verbal prompts telling the player what to do needed to be implemented. 

Future work, outside the scope of our summer project, is required to verify that our modifications to the gaming environment enhanced the playability and learning objectives through Games User Research methods. We suggest using a research approach with two groups of test subjects, each exploring a different version of the game. Game playability should be assessed using a mixed approach with both quantitative and qualitative measurements that investigate the effects of playing the game on a player's understanding of blood cell physiology.

Future development, should also look to refine elements within the game, including expansion of the axis of movement on the microscope slide and in the blood vessel, which were restrained due to concerns about motion sickness in a VR state and challenges with physics. In this development, fine-tuning, including changes to frame rate and expansion to other VR devices, should also be implemented to make the experience more accessible for patients. We also aim to apply our iterative game design process to the remainder of the LeukemiaSIM mini-games to make a cohesive gaming experience. Once developed, the full experience should touch on acute lymphoblastic leukemia pathophysiology and physiology using realistic science concepts developed to enhance player engagement in VR.

\section{Conclusion}

As we move forward, it is important to reflect on the fact that virtual learning is independent of conventional learning and needs to manifest itself to reflect the virtual space. Instead of implementing environments that mirror reality, game developers should think outside of the box and create spaces that are otherwise inaccessible. These spaces should reflect scientific concepts to enhance hands-on-learning, but should also incorporate learning environments that are inaccessible outside of the virtual space. Every VR headset is a metaphorical Magic School Bus.

In our refinement of the Red Blood Cell Oxygenation Game and the Blood Cell Interaction Zone, we redesigned the virtual space to better reflect science and enable hands-on-learning by utilizing player engagement techniques. We believe that our new games and simulations will provide memorable learning experiences for acute lymphoblastic leukemia patients and their families, as they explore scientifically accurate representations of blood cells in a virtual world.

\bibliographystyle{unsrt} 
\bibliography{references}

\begin{thebibliography}{10}

\bibitem{sanchez_playability_2012}
José Luis~González Sánchez, Francisco Luis~Gutiérrez Vela,
  Francisco~Montero Simarro, and Natalia Padilla-Zea.
\newblock Playability: analysing user experience in video games.
\newblock {\em Behaviour \& Information Technology}, 31(10):1033--1054, October
  2012.

\bibitem{sherman_chapter_2019}
William~R. Sherman and Alan~B. Craig.
\newblock Chapter 1 - {Introduction} to {Virtual} {Reality}.
\newblock In William~R. Sherman and Alan~B. Craig, editors, {\em Understanding
  {Virtual} {Reality} ({Second} {Edition})}, The {Morgan} {Kaufmann} {Series}
  in {Computer} {Graphics}, pages 4--58. Morgan Kaufmann, Boston, January 2019.

\bibitem{bowman_educational_1999}
Doug~A. Bowman, Larry~F. Hodges, Don Allison, and Jean Wineman.
\newblock The {Educational} {Value} of an {Information}-{Rich} {Virtual}
  {Environment}.
\newblock {\em Presence: Teleoperators and Virtual Environments},
  8(3):317--331, June 1999.
\newblock Publisher: MIT Press.

\bibitem{pantelidis_reasons_2010}
Veronica~S. Pantelidis.
\newblock Reasons to {Use} {Virtual} {Reality} in {Education} and {Training}
  {Courses} and a {Model} to {Determine} {When} to {Use} {Virtual} {Reality}.
\newblock {\em Themes in Science and Technology Education}, 2(1-2):59--70,
  October 2010.
\newblock Number: 1-2.

\bibitem{yavrucuk_low_2011}
Ilkay Yavrucuk, Eser Kubali, and Onur Tarimci.
\newblock A low cost flight simulator using virtual reality tools.
\newblock {\em IEEE Aerospace and Electronic Systems Magazine}, 26(4):10--14,
  April 2011.
\newblock Conference Name: IEEE Aerospace and Electronic Systems Magazine.

\bibitem{loftin_applying_1993}
R.~Loftin, M.~Engleberg, and R.~Benedetti.
\newblock Applying virtual reality in education: {A} prototypical virtual
  physics laboratory.
\newblock In {\em Proceedings of 1993 {IEEE} {Research} {Properties} in
  {Virtual} {Reality} {Symposium}}, pages 67--74, San Jose, California, USA,
  1993. IEEE.

\bibitem{johnson_augmenting_2002}
A.~Johnson, T.~Moher, Y.J. Cho, Y.J. Lin, D.~Haas, and J.~Kim.
\newblock Augmenting elementary school education with {VR}.
\newblock {\em IEEE Computer Graphics and Applications}, 22(2):6--9, March
  2002.
\newblock Conference Name: IEEE Computer Graphics and Applications.

\bibitem{huang_investigating_2010}
Hsiu-Mei Huang, Ulrich Rauch, and Shu-Sheng Liaw.
\newblock Investigating learners’ attitudes toward virtual reality learning
  environments: {Based} on a constructivist approach.
\newblock {\em Computers \& Education}, 55(3):1171--1182, November 2010.

\bibitem{jimenez_patient_2018}
Yobelli~A. Jimenez, Steven Cumming, Wei Wang, Kirsty Stuart, David~I. Thwaites,
  and Sarah~J. Lewis.
\newblock Patient education using virtual reality increases knowledge and
  positive experience for breast cancer patients undergoing radiation therapy.
\newblock {\em Supportive Care in Cancer}, 26(8):2879--2888, August 2018.

\bibitem{pandrangi_application_2019}
Vivek~C Pandrangi, Brandon Gaston, Nital~P Appelbaum, Francisco~C
  Albuquerque~Jr, Mark~M Levy, and Robert~A Larson.
\newblock The {Application} of {Virtual} {Reality} in {Patient} {Education}.
\newblock {\em Annals of Vascular Surgery}, 59:184--189, August 2019.
\newblock Publisher: Elsevier.

\bibitem{kuehn_virtual_2018}
Bridget~M. Kuehn.
\newblock Virtual and {Augmented} {Reality} {Put} a {Twist} on {Medical}
  {Education}.
\newblock {\em JAMA}, 319(8):756--758, February 2018.

\bibitem{reznek_virtual_2002}
Martin Reznek, Phillip Harter, and Thomas Krummel.
\newblock Virtual reality and simulation: training the future emergency
  physician.
\newblock {\em Academic Emergency Medicine: Official Journal of the Society for
  Academic Emergency Medicine}, 9(1):78--87, January 2002.

\bibitem{hoffman_use_2000}
H.~G. Hoffman, D.~R. Patterson, and G.~J. Carrougher.
\newblock Use of virtual reality for adjunctive treatment of adult burn pain
  during physical therapy: a controlled study.
\newblock {\em The Clinical Journal of Pain}, 16(3):244--250, September 2000.

\bibitem{bridge_development_2007}
P.~Bridge, R.~M. Appleyard, J.~W. Ward, R.~Philips, and A.~W. Beavis.
\newblock The development and evaluation of a virtual radiotherapy treatment
  machine using an immersive visualisation environment.
\newblock {\em Computers \& Education}, 49(2):481--494, September 2007.
\newblock Publisher: Elsevier Ltd.

\bibitem{birnie_usability_2018}
Kathryn~A. Birnie, Yalinie Kulandaivelu, Lindsay Jibb, Petra Hroch, Karyn
  Positano, Simon Robertson, Fiona Campbell, Oussama Abla, and Jennifer
  Stinson.
\newblock Usability {Testing} of an {Interactive} {Virtual} {Reality}
  {Distraction} {Intervention} to {Reduce} {Procedural} {Pain} in {Children}
  and {Adolescents} {With} {Cancer} :.
\newblock {\em Journal of Pediatric Oncology Nursing}, 35(6):406--416, June
  2018.
\newblock Publisher: SAGE PublicationsSage CA: Los Angeles, CA.

\bibitem{hoffman_virtual_2011}
Hunter~G. Hoffman, Gloria~T. Chambers, Walter~J. Meyer, Lisa~L. Arceneaux,
  William~J. Russell, Eric~J. Seibel, Todd~L. Richards, Sam~R. Sharar, and
  David~R. Patterson.
\newblock Virtual reality as an adjunctive non-pharmacologic analgesic for
  acute burn pain during medical procedures.
\newblock {\em Annals of Behavioral Medicine: A Publication of the Society of
  Behavioral Medicine}, 41(2):183--191, April 2011.

\bibitem{atzori_virtual_2018}
Barbara Atzori, Hunter~G. Hoffman, Laura Vagnoli, David~R. Patterson, Wadee
  Alhalabi, Andrea Messeri, and Rosapia Lauro~Grotto.
\newblock Virtual {Reality} {Analgesia} {During} {Venipuncture} in {Pediatric}
  {Patients} {With} {Onco}-{Hematological} {Diseases}.
\newblock {\em Frontiers in Psychology}, 9:2508, 2018.

\bibitem{drachen_gur_2018}
Anders Drachen, Pejman Mizra-Babei, and Lennart~E. Nacke.
\newblock {\em {GUR} {Book} – {The} {Games} {User} {Research} {Book}}.
\newblock Oxford University Press, New York, NY, USA, 2018.

\bibitem{livingston_post-launch_2018}
Ian Livingston.
\newblock {\em Post-launch in {Games} {User} {Research}}, volume~1.
\newblock Oxford University Press, March 2018.

\bibitem{cole_magic_2020}
Joanna Cole and Bruce Degen.
\newblock The {Magic} {School} {Bus} {\textbar} {Books}, {Experiments},
  {Printables}, {Apps} {\textbar} {Scholastic} {Kids}, 2020.

\bibitem{fullerton_game_2013}
Tracy Fullerton.
\newblock {\em Game {Design} {Workshop}: a {Playcentric} {Approach} to
  {Creating} {Innovative} {Games}, {Third} {Edition}.}
\newblock CRC Press, Oakville, 2013.
\newblock OCLC: 967110587.

\bibitem{flanagan_critical_2009}
Mary Flanagan.
\newblock {\em Critical play radical game design}.
\newblock MIT Press, Cambridge, MA, 2009.
\newblock OCLC: 1162028111.

\bibitem{joyce_10_2019}
Alita Joyce.
\newblock 10 {Usability} {Heuristics} {Applied} to {Video} {Games}, May 2019.
\newblock Library Catalog: www.nngroup.com.

\bibitem{giannakos_enjoy_2013}
Michail~N. Giannakos.
\newblock Enjoy and learn with educational games: {Examining} factors affecting
  learning performance.
\newblock {\em Computers \& Education}, 68:429--439, October 2013.

\bibitem{mcluhan_war_1968}
Marshall McLuhan and Quentin. Fiore.
\newblock {\em War and peace in the global village; an inventory of some of the
  current spastic situations that could be eliminated by more feedforward}.
\newblock McGraw-Hill, New York, 1968.

\end{thebibliography}

\end{document}